\pdfoutput=1  %arXiv addition for pdflatex, permits no file extensions.
\documentclass[]{article}  
\usepackage{url,float}
\usepackage{graphicx}
\usepackage{amsmath}
\usepackage{amsfonts}
\usepackage{amssymb}
\usepackage{latexsym}

\newcommand{\hide}[1]{}

\newcommand{\ABox}{
\raisebox{3pt}{\framebox[6pt]{\rule{6pt}{0pt}}}
}
\newenvironment{proof}{{\bf Proof:}}{\hfill\ABox}

\newtheorem{theorem}{{\bf Theorem}}

\newtheorem{lemma}{Lemma}
\newcommand{\lemlab}[1]{\label{lemma:#1}}
\newcommand{\thmlab}[1]{\label{thm:#1}}

\newcommand{\figlab}[1]{\label{fig:#1}}
\newcommand{\seclab}[1]{\label{sec:#1}}

\newcommand{\thmref}[1]{\ref{thm:#1}}

\newcommand{\figref}[1]{\ref{fig:#1}}

\def\C{{\mathcal C}}
\def\A{{\mathcal A}}
\def\R{{\mathbb{R}}}
\def\S{{\mathbb{S}}}

\title{%
Convex Polyhedra Realizing Given Face Areas
} %title

\author{%
Joseph O'Rourke%
    \thanks{Department of Computer Science, Smith College, Northampton, MA
      01063, USA.
      \protect\url{orourke@cs.smith.edu}.}
}%author

\begin{document}
\maketitle

\begin{abstract}
Given $n \ge4$ positive real numbers, we prove in this note that they are the face
areas of a convex polyhedron if and only if the largest
number is not more than the sum of the others.
\end{abstract}

\section{Introduction}
\seclab{Introduction}
% ~\cite{a-cp-05}.
% ~\cite{do-gfalop-07}
% ~\cite{o-cgc-98}

Let $\A=(A_1,A_2,\ldots,A_n)$ be a vector of $n$ positive real numbers
sorted so that $A_i \ge A_{i+1}$.
The question we address is in this note is:
\begin{quotation}
\noindent
When does $\A$ represent the face areas of a convex polyhedron in $\R^3$?
\end{quotation}
For example, suppose $\A=(100,1,1,1)$. It is clear there is no
tetrahedron
realizing these areas, because the face of area 100 is too large
to be ``covered'' by the three faces of area 1.
So $A_1 \le \sum_{i>1} A_i$ is an obvious necessary condition.
The main result of this note is that this is also a sufficient
condition.
Enroute to establishing this we connect the question to robot arm
linkages and to 3D polygons.

The main tool we use is Minkowski's 1911 theorem.
Here is a version
from Alexandrov,
who devotes an entire chapter to the theorem and 
variations in his book~\cite[Chap.~7, p.~311ff]{a-cp-05}.

\begin{theorem}[Minkowski (a)]
Let $A_i$ be positive faces areas and $n_i$ distinct, noncoplanar unit face normals,
$i=1,\ldots,n$.
Then if $\sum_i A_i  n_i = 0$, there is a closed polyhedron
whose faces areas uniquely realize those areas and normals.
\thmlab{Mink1}
\end{theorem}
\noindent
Here, uniqueness is up to translation.

In our situation, we are given the areas $A_i$, and the task is to 
determine if there exist normals $n_i$ that satisfy Minkowski's
theorem.
Although this superficially may seem like a complex problem, we will
see it has a simple solution.
Although I have not been able to find this result in the literature,
it seems likely that it is known, because the proof is not difficult.

%\subsection{Gr\"{u}nbaum}

It will be more convenient for our purposes
to follow Gr\"{u}nbaum's (equivalent) formulation
of Minkowski's theorem~\cite[p.~332]{g-cp-03}
phrased in terms of ``fully equilibrated'' vectors.
Vectors are
\emph{equilibrated}
if they sum to zero and no two are positively proportional.
They are 
\emph{fully equilibrated}
in $\R^k$ if they in addition span $\R^k$.

\addtocounter{theorem}{-1}

\begin{theorem}[Minkowski (b)]
Let $v_i = A_i n_i$ be vectors whose lengths are $A_i$, $|v_i| = A_i$,
and whose directions are unit normal vectors $n_i$,
$i=1,\ldots,n$.
Then if the vectors are fully equilibrated in $\R^3$,
there is a unique closed polyhedron $P$
with faces areas $A_i$ and normal vectors $n_i$.
\thmlab{Mink2}
\end{theorem}

Note that for $n$ vectors to be  fully equilibrated in $\R^3$, 
$n$ must be at least 4:
It requires 3 vectors to span $\R^3$, but any three vectors
that sum to zero form a triangle and so lie in a plane.
Thus 3 vectors cannot both be equilibrated and span $\R^3$.
Four clearly suffices: $P$ is a tetrahedron.

\section{Main Result \& Proof}
\seclab{Proof}

\begin{theorem}
If $A_1,A_2,\ldots,A_n$ are positive real numbers with $n \ge 4$ and
$A_i \ge A_{i+1}$, then there is a closed convex polyhedron $P$
with these $A_i$ as its face areas if and only if
$A_1 \le \sum_{i>1} A_i$.
When equality holds,
we permit the polyhedron to be \emph{flat}, i.e.,
its faces tessellate and doubly cover a planar convex polygon.
\thmlab{Main}
\end{theorem}
%\begin{proof}

\subsection{Flat Polyhedra}
Let $F_i$ be the face whose area is $A_i$.
As previously mentioned, the condition is necessary,
because if $A_1 > \sum_{i>1} A_i$,
then the $F_1$ face cannot be covered by all the others, so it is not
possible to form a closed surface.
In the case of equality, $A_1 = \sum_{i>1} A_i$,
the areas are realized by a flat polyhedron that can be
constructed as follows.
For $F_1$, select a square of side length $\sqrt{A_1}$.
(Any other convex polygon would serve as well.)
This becomes one side of $P$.
For the other side, partition $F_1$ into strips of
width $A_i / \sqrt{A_1}$, so that each strip has area $A_i$,
and serves as $F_i$.

Henceforth assume $A_1$ is strictly smaller than the sum of the other
areas.
In this circumstance we can always obtain a non-flat polyhedron $P$,
with none of the faces $F_i$ coplanar.
% Let $k \ge 1$ be the largest index such that
% $$A_1 + \cdots + A_k \le A_{k+1} + \cdots + A_n \;.$$
% In the case of equality, we repeat the previous argument,
% this time partitioning a square into $k$ strips for $F_1,\ldots,F_k$,
% which becomes one side of a flat $P$, and
% partitioning the other side into $n-k$ strips for
% $F_{k+1},\ldots,F_n$.
% Assume now that we have strict inequality by a gap of $\d$:
% $$\sum_{i=1}^k A_i + \d = \sum_{i=k+1}^n A_i \;.$$
%\end{proof}

\subsection{Robot Arms}
Let $\C$ be a polygonal chain (sometimes called a ``robot arm'' or
just ``arm'')
whose $n$ links have lengths $A_1, A_2, ..., A_n$.
Then it is a corollary to a theorem of Hopcroft, Joseph, and Whitesides~\cite{hjw-mp2dl-84}
that the chain can close (i.e., the ``hand'' can touch the
``shoulder'')
iff the longest link is not longer than all the other links together.
See, e.g.,~\cite[Thm.~8.6.3, p.~326]{o-cgc-98}
or ~\cite[Thm.~5.1.2, p.~61]{do-gfalop-07}.
The similarity to the statement of Theorem~\thmref{Main} should be
evident.
Our plan is to form a chain $\C$ from fully equilibrated vectors,
and apply Minkowski's theorem, Theorem~\thmref{Mink2}(b).
This can be accomplished in three stages:
(1)~arrange that the vectors sum to zero,
(2)~ensure that none is a positive multiple of another,
and (3)~ensure that they span $\R^3$.
We know from the robot-arm theorem that~(1) is achievable, but
that theorem is an existence theorem.

Satisfying~(1) can be accomplished with
the ``Two Kinks'' 
theorem~\cite[Thm.~8.6.5, p.~329]{o-cgc-98}~\cite[Thm.~5.1.4,
p.~62]{do-gfalop-07},
which would result in the links arranged to form a
triangle (and so summing to 0), with (in general) many vectors aligned.  Although it is then
not so difficult to break all the alignments and achieve~(2),
instead we opt for a method that
achieves~(1) and~(2) simultaneously.

\subsection{Cyclic Polygon}
Lay out all the links in a straight line of length $\sum_i A_i$.
View the links as inscribed in a circle of infinite radius.
Now imagine shrinking the radius from $R=\infty$ down toward $R=0$,
maintaining the chain inscribed at all times.
Knowing from the Hopcroft et al. theorem that the chain can close,
at some radius $R$ it just barely closes up, and we have a cyclic polygon $\C$.
See Figure~\figref{Cyclic}.
%%%%%%%%%%%%%%%%%%%%%%%%%%%%%%%%%Figure Begin
\begin{figure}[htbp]
\centering
\includegraphics[width=\linewidth]{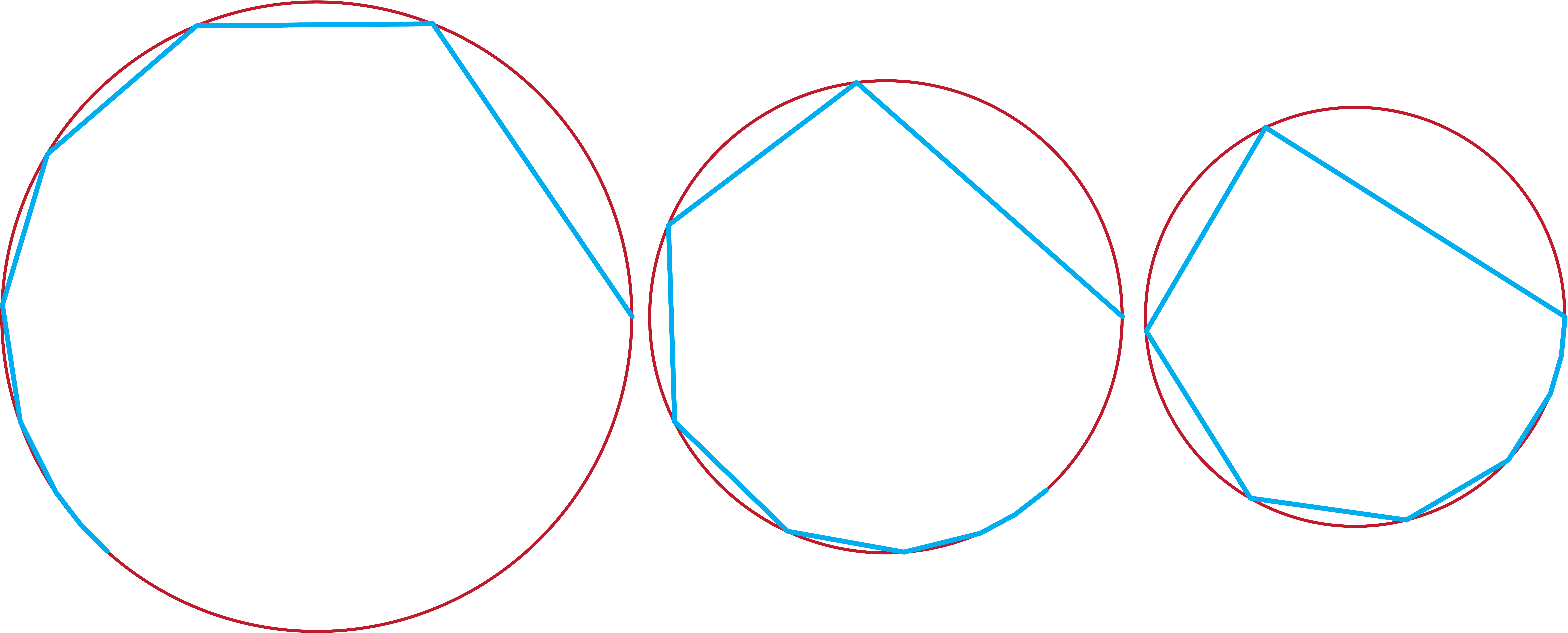}
%\fbox{Cyclic}
\caption{Links of lengths $(9, 6, 5, 4, 3, 2, 1, 1)$
inscribed in circles of radius $R=8$, $R=6$, and closing at $R \approx
5.325$.}
\figlab{Cyclic}
\end{figure}
%%%%%%%%%%%%%%%%%%%%%%%%%%%%%%%%%Figure End

Let $v_1, v_2, ..., v_n$ be the vectors comprising this cyclic $\C$.
They form a planar convex polygon connected head to tail,
with no two vectors aligned.
These vectors are therefore equilibrated, but they do not span $\R^3$,
so they are not fully equilibrated in $\R^3$ in Gr\"{u}nbaum's terminology.

\subsection{Spanning $\R^3$}
Arranging the vectors to span $\R^3$ is easily accomplished, in many
ways.
Here is one.  
Let $a$ be the base of $v_1$, and select some $k$ in $(2,3,\ldots,n-2)$
(so for $n=4$ we must have $k=2$).
Let $b$ be the head of $v_k$ around the convex polygon.
See Figure~\figref{ConvexSpan}.
%%%%%%%%%%%%%%%%%%%%%%%%%%%%%%%%%Figure Begin
\begin{figure}[htbp]
\centering
\includegraphics[width=0.9\linewidth]{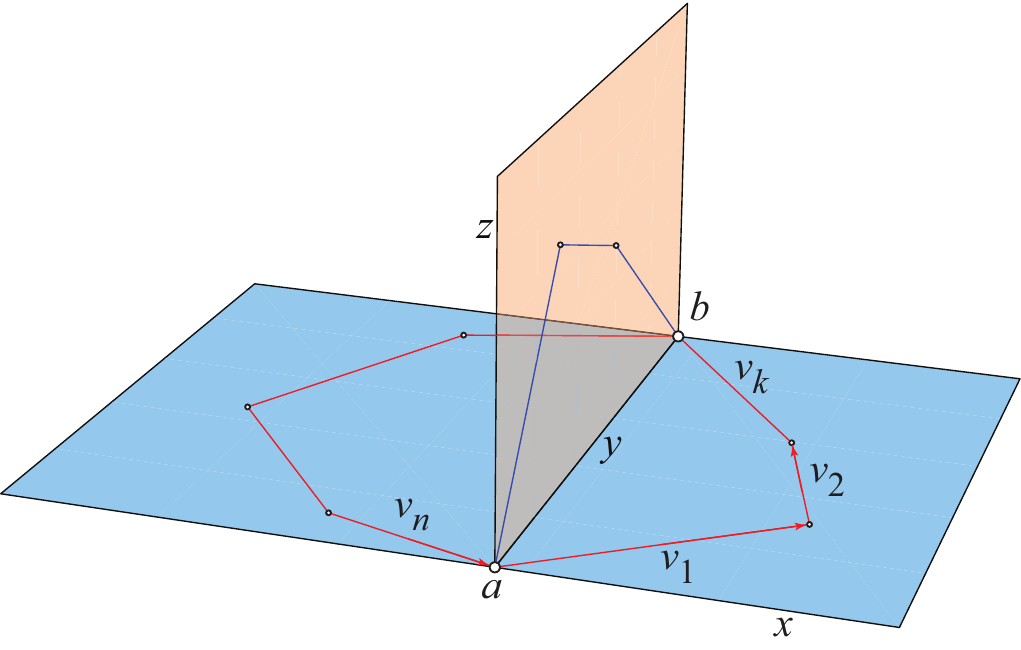}
%\fbox{ConvexSpan}
\caption{Rotating $v_1,\ldots,v_k$ to the $yz$-plane.}
\figlab{ConvexSpan}
\end{figure}
%%%%%%%%%%%%%%%%%%%%%%%%%%%%%%%%%Figure End
Rotate the portion (we'll call it ``half'') of the convex polygon including $v_1,\ldots,v_k$
around the line through $ab$ until that portion lies in a vertical
plane, say, the $yz$-plane.
Now half the vectors lie in this $yz$-plane, and the other
half lie in the $xy$-plane.
Each half contains at least two vectors by our choice of $k$.
Thus the vectors in the $xy$-plane span that plane, and the vectors
in the $yz$-plane span that plane.
Consequently, together they span $\R^3$.

\subsection{Proof Completion}
Finally we may apply Minkowski's Theorem, Theorem~\thmref{Mink2}(b)
to conclude that there is a closed, convex polyhedron $P$
whose face areas are the lengths of the vectors $v_i$,
$|v_i|=A_i$.

\section{Discussion}

\begin{enumerate}

\item
Theorem~\thmref{Main} still holds for $n=2$, when
two given areas $A_1=A_2$ are realized by a flat two-face polyhedron.
But it does not hold for $n=3$ (except in the case of equality,
$A_1=A_2+A_3$, by a flat polyhedron).
For $n=3$, the theorem condition is the
triangle inequality, and one obtains a triangle with side lengths the
three given ``areas.''
One could construct an infinite triangular prism with these face area ratios.

\item
The value of $R$ that closes $\C$ to a cyclic polygon satisfies this
equation,
where each term in the sum is the angle at the circle center subtended
by $A_i$:
\begin{equation}
\sum_i 2 \sin^{-1} \left( \frac{1}{2} A_i / R \right) = 2 \pi
\end{equation}
$R$ can easily be computed numerically;
I cannot see a more direct computation.

\item
An artifact of the method used to ensure the vectors span $\R^3$ is
that
all the normal vectors of the polyhedron $P$ lie in two orthogonal
planes.
The resulting polyhedra all have roughly the shape of the intersection
of
two half-cylinders, as depicted in
Figure~\figref{CylinderCylinder}.
%%%%%%%%%%%%%%%%%%%%%%%%%%%%%%%%%Figure Begin
\begin{figure}[htbp]
\centering
\includegraphics[width=0.75\linewidth]{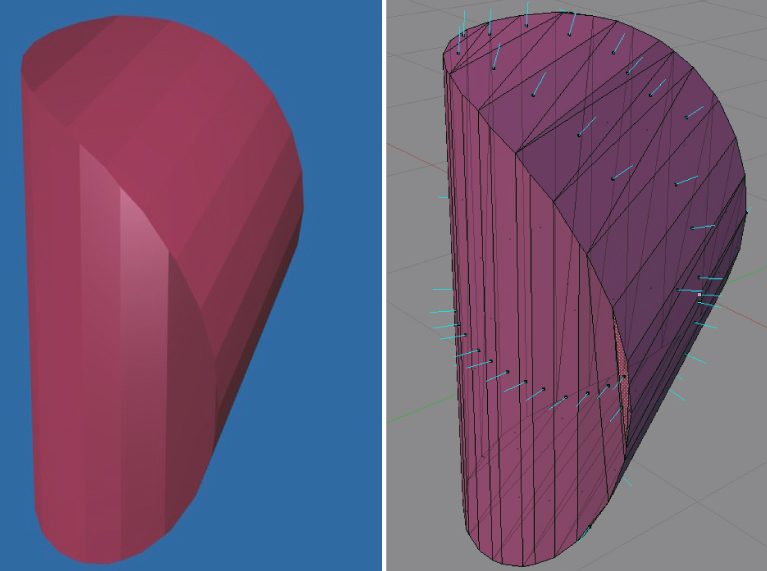}
%\fbox{CylinderCylinder}
\caption{Intersection of two half-cylinders. Right image shows normal vectors.}
\figlab{CylinderCylinder}
\end{figure}
%%%%%%%%%%%%%%%%%%%%%%%%%%%%%%%%%Figure End
Instead of spinning half the vectors about $ab$, we could spin each
pair
of successive vectors $(v_i, v_{i+1})$ about the line containing their
sum $v_i + v_{i+1}$. 
Choosing the spins independently would result in
a more ``balanced'' collection of normals $n_i$.

\item
Along these lines, it would be pleasing to have a natural
candidate for a ``canonical polyhedron'' that realizes given areas.

\item
In the 1980s, Little described a constrained optimization procedure that could
compute the polyhedron guaranteed by Minkowski's
theorem~\cite{l-egimv-85}.
Unfortunately I know of no modern implementation.

\item 
Minkowski's theorem generalizes to $\R^d$, for any $d \ge 2$.
It is clear that,
given facet volumes $V_1, \ldots, V_n$, 
the analog of Theorem~\thmref{Main} 
holds.
One only need ensure that the vectors span $\R^d$
by choosing spins of adjacent vectors about ``planes of rotation''
that rotate the vectors to span the dimensions beyond $R^3$.
I will leave this as a claim requiring further work to justify precisely.

\item
The ``configuration space'' (or ``moduli space'') of all the polyhedra that realize
a given list of areas is the same as the configuration space
of a 3D polygon (closed polygonal chain) with those areas as edge lengths.
This configuration space is known to be connected,
by a result of Lenhart and Whitesides~\cite{lw-rcpce-95}
(see also~\cite[Thm.~5.1.9, p.~67]{do-gfalop-07}).
The space is well studied, e.g.~\cite{pt-gps-07},~\cite{m-nmsp-08}.
For a quadrilateral, the space is topologically a sphere $\S^2$,
but its structure is more complicated for arbitrary $n$-gons.

% These configuration spaces when the polygon is restricted to be planar
% have been especially well-studied:
% \cite{km-mspep-95} and~\cite{sv-sppmt-05}.

\end{enumerate}

% \paragraph{Acknowledgments.}

\bibliographystyle{alpha}
\bibliography{/Users/orourke/bib/geom/geom}
\end{document}